\documentclass[12pt]{iopart}

\usepackage[dvips]{graphicx}
\usepackage{psfrag}

\begin{document}

\title[disordered bosons in a lattice]
{Mean-field phase diagram of disordered bosons in a lattice at non-zero temperature}

\author{K V Krutitsky, A Pelster and R Graham}

\address{Fachbereich Physik der Universit{\"a}t Duisburg-Essen, Campus Essen,
Universit{\"a}tsstr.~5, 45117 Essen, Germany}

\ead{kostya@theo-phys.uni-essen.de}

\begin{abstract}
Bosons in a periodic lattice with on-site disorder at low but non-zero temperature
are considered within a mean-field theory. 
The criteria used for the definition of the superfluid, Mott insulator and Bose glass
are analysed. Since the compressibility does never vanish at non-zero temperature,
it can not be used as a general criterium. We show that the phases are 
unambiguously distinguished by the superfluid density and the density of states
of the low-energy exitations. The phase diagram of the system is calculated.
It is shown that even a tiny temperature leads to a significant
shift of the boundary between the Bose glass and superfluid.
\end{abstract}



\submitto{\NJP}


\section{Introduction}

The remarkable experimental control over ultracold atomic gases in optical lattices
acquired in the last couple of years~\cite{Bloch1,Zwerger,Esslinger,Zoller,Bloch2,Review} 
has opened up completely new lines
of interest in the field of Bose-Einstein condensation. One of these are ultracold
atoms in optical lattices with disorder which can be created by several methods.
One of the possibilities is to use a laser speckle field~\cite{Inguscio1,Inguscio2,Aspect}.
An alternative way
of creating a disorder potential is the introduction of a tiny fraction of a second
atomic species 
which are strongly localized on random sites~\cite{Vignolo,Castin,Sengstock}. 
Tuning to a Feshbach resonance, these random
scatterers can even make the disorder very strong. 
The random potentials for atoms can be also created via the spatial fluctuations 
of the electric currents generating the magnetic wire traps~\cite{Schmiedmayer,GWSS05,GWSS06}
or with the aid of the incommensurate lattices~\cite{Lew03,Santos05}.
Disordered lattices for ultracold rubidium atoms have been recently created
by superimposing a regular periodic optical potential on the speckle field~\cite{Schulte}
and on the incommensurate lattice~\cite{NEU}. 

These very recent experimental developments prompt the theoretical question what the
behavior of ultracold atomic gases in lattice potentials with disorder might be.
Typical for the modern area of Bose-Einstein condensation (BEC) in particular and 
ultracold atomic gases in general is that such basic theoretical questions have been
asked before in a different context. 
Indeed, there is an enormous body of literature on ultracold
Fermi gases in disordered lattices referring, first of all, to electrons in amorphous
solids, either in the normal or in the superconducting state~\cite{Kirkpatrick}. The possibility
to study fermionic atomic gases from this point of view is a very active and
interesting field with which we shall not deal here, however. Suffice it to say
that completely new questions appear like the 
BEC-BCS crossover~\cite{BECBCS1,BECBCS2}. The influence
of disorder on this problem is still completely unknown.

Here, we shall focus
exclusively on bosons in potentials with disorder. This is also known as the 'dirty
boson problem'. It first came up in the pre-BEC area in the context of experimental
investigations of superfluidity of $^{4}$He in the random pores of Vycor. The surprising
finding that, for sufficiently low coverage of the pores, the $^{4}$He superfluidity would
disappear even in an extrapolation to zero temperature~\cite{Reppy1,Reppy2,Reppy3} prompted
many theoretical studies. These were based on Hartree-Fock 
theory~\cite{Anderson,Ordnungsparameter},
generalizations of the Bogoliubov and the Beliaev theory~\cite{AGD}
for random potentials~\cite{Gunn,Zhang1,Zhang2,Huang,Giorgini,SR,Kobayashi,Vinokur}, 
field-theoretical considerations~\cite{Lee,Fisher}, 
mean-field theory~\cite{Fisher,Pazmandi},
renormalization group theory~\cite{Sing,Pai,Svistunov,PZ,Rapsch} and
quantum Monte Carlo simulations~\cite{Krauth,Batrouni,WSGY,KR,KR2,Park,Kim,LeeCha,PS,Hitchcock}
as well as numerical diagonalizations~\cite{Runge}. 
The consensus which developed from these 
studies is that in a disordered lattice at temperature $T=0$ two new
phases of bosons may exist besides the superfluid phase which is theoretically defined
by the presence of the off-diagonal long-range order: One is the Mott-insulator phase,
which only exists at commensurate fillings of the lattice, and is distinguished by the absence
of the off-diagonal long-range order, a non-zero energy gap and a vanishing compressibility.
The other is the Bose-glass phase, which is distinguished again by the absence
of the off-diagonal long-range order, a non-vanishing density of states at zero energy, and
a non-zero compressibility. A more recent suggestion towards identifying the Bose-glass phase
has been made in Ref.~\cite{Davidson}.

While these operational definitions of a Mott-insulator
phase and a Bose-glass phase at temperature
$T=0$ are precise and clear-cut, they run into the obvious difficulty that experiments 
and quantum Monte Carlo simulations are
never performed at $T=0$. It is therefore necessary to examine the extent to which
these or similar definitions can be applied at least at small non-zero temperature. This
is the goal of the present paper. In order to achieve our goal we have to investigate
the low-lying states of a suitable model of strongly interacting bosons
in a lattice with disorder. We shall choose for this purpose a Bose-Hubbard
model with on-site disorder of bounded variation. For our purpose of defining
and distinguishing the various phases at non-zero temperature it is sufficient to analyse
the basic model within a mean-field theory which allows to detect all the phases
and to see what happens if the system parameters are changed, although it does not
provide precise conditions for various phase transitions encountered in the system.

The paper is organized as follows. In Section 2 the model and the mean-field approach
to its analysis are defined. Then, in Section 3,
the phase boundary between the superfluid and the two
non-superfluid phases is derived from the condition of vanishing off-diagonal long-range
order. This is first done for the pure case without disorder, reproducing a well-known
result, and then generalizing it to the case with disorder. In Section 4 follows
the definition of the Bose-glass and Mott-insulator phases at non-zero temperature and the 
examination of the phase boundary between them. A common criterion for the distinction
of these two phases at $T=0$ is the compressibility. How well this quantity serves
this purpose at finite temperature is therefore examined in Section 5. In Section 6 the paper
ends with some final conclusions.

\section{Hamiltonian}

We consider a system of spinless bosons in a homogeneous infinitely extended lattice
of dimension $d=1,2,3$ described by the Bose-Hubbard Hamiltonian
(in units of $\hbar=1$)
\begin{equation}
H_{BH}
=
-J
\sum_{<\bi{i},\bi{j}>}
a_\bi{i}^\dagger
a_\bi{j}
+
\frac{U}{2}
\sum_\bi{i}
a_\bi{i}^\dagger
a_\bi{i}^\dagger
a_\bi{i}
a_\bi{i}
-
\sum_\bi{i}
\left(
    \mu
    +
    \epsilon_\bi{i}
\right)
a_\bi{i}^\dagger
a_\bi{i}
\;,
\end{equation}
where $J$ is the tunneling matrix element, $U$ is the on-site interaction constant,
and $\mu$ is the chemical potential.
In this work we assume that the random on-site energies $\epsilon_\bi{i}$ 
at different sites are uncorrelated 
and equally distributed with the probability density $p(\epsilon)$.

We introduce the superfluid order parameter $\psi=\overline{\langle a_\bi{i}\rangle}$,
where
$
 \langle\dots\rangle
 =
 \Tr
 \left[
     \dots \exp(-\beta H)
 \right]
 /
 \Tr
 \left[
     \exp(-\beta H)
 \right]
$
and
$
 \overline{\phantom{|}\dots\phantom{|}}
 =
 \Pi_\bi{i}
 \int_{-\infty}^{+\infty}
 \dots p(\epsilon_\bi{i}) d\epsilon_\bi{i}
$
denote quantum-mechanical and disorder averaging, respectively, and $\beta=1/(kT)$. 
Making use of the decoupling mean-field approximation in the
hopping term~\cite{SKPR,Sachdev,Oosten} which is valid for sufficiently high-dimensional
systems, we obtain the following on-site Hamiltonian
\begin{equation}
\label{hmf}
H
=
-2dJ
\left(
    \psi a^\dagger
    +
    \psi^* a
\right)
+
2dJ
\left|
    \psi
\right|^2
+
\frac{U}{2}
a^\dagger a^\dagger a \; a
-
\left(
    \mu
    +
    \epsilon
\right)
a^\dagger
a
\;,
\end{equation}
where we have omitted the site index.

The phase diagram of the system can be obtained in the following manner.
First of all one has to calculate the disorder-averaged free energy of the system 
$\overline{F(\psi)}$
corresponding to the Hamiltonian~(\ref{hmf}).
Then minimizing it with respect to $\psi$ to determine $\psi=\psi_m$ 
one can distinguish the superfluid ($\psi_m\neq0$)
and non-superfluid ($\psi_m=0$) regions of the parameter space.
Applying a small phase gradient to the atomic matter-field operator and calculating
the corresponding correction to the free energy in a manner similar to 
Ref.~\cite{RB03}, one can show that the superfluid density following from
the Hamiltonian~(\ref{hmf}) equals $\left|\psi_m\right|^2$.
In the non-superfluid region, one has to work out the disorder average of 
the static superfluid susceptibility $\overline{\chi}$
or the density of states $\overline{\rho(\omega)}$ for the single-particle 
excitations~\cite{Fisher}.
In the region where $\overline{\rho(\omega)}=0$ in the interval $0\leq\omega<\omega_g$
we have, by the definition we apply, the Mott-insulator phase with the energy gap $\omega_g$.
On the other hand, again by definition, the Bose-glass phase occurs when 
$\lim_{\omega\to0}\overline{\rho(\omega)}\neq0$ which corresponds to a
divergent superfluid susceptibility~\cite{Fisher}.

The form of the mean-field Hamiltonian (\ref{hmf}) implies that, in our
approximation, the properties of the
Mott-insulator phase as well as the Bose-glass phase, where $\psi$ vanishes,
do not depend on the tunneling matrix element $J$. This is consistent with the
fact that the boundary between these phases occurs only for small values of $J$. 
The transition to superfluidity, where $\psi_m\ne0$ starts to appear, does depend on $J$
also in our approximation.

\section{Boundary between the superfluid and non-superfluid phases}

In order to calculate the free energy of the system, one has to solve the eigenvalue problem
for the Hamiltonian (\ref{hmf}). This can be done exactly by means of numerical calculations.
However, the boundary between the superfluid and non-superfluid phases can also be determined
with high accuracy treating the first term in the Hamiltonian (\ref{hmf}) as a perturbation.
The free energy can only depend on $\left|\psi\right|^2$ since a change of the phase in $\psi$
can be undone by the unitary transformation 
$a\to\exp(-i\varphi)a$, $a^\dagger\to\exp(i\varphi)a^\dagger$.
Indeed, the calculations show that the result has the following structure:
\begin{equation}
\overline{F(\psi)}
=
\overline{a_0}
+
\overline{a_2}
\left|
    \psi
\right|^2
+
\overline{a_4}
\left|
    \psi
\right|^4
+
\dots
\end{equation}
The explicit form of $a_4$ as well as $a_0$ and $a_2$ for $T=0$ was obtained in 
Ref.~\cite{Oosten}.
The generalization to $T\neq0$ and the average over disorder needed here is straightforward.
Since $\overline{a_4}$ turns out to be always positive and $\overline{a_2}$ can be 
either positive
or negative, the superfluid/non-superfluid transition is of second order.
The equation $\overline{a_2}=0$ determines the phase boundary which is given by
\begin{eqnarray}
\label{phb}
&&
\int_{-\infty}^{+\infty}
\frac{d \mu'  \; p(\mu - \mu')}{Z_0(\mu')}
\sum_{m=0}^\infty
\left[
    \frac{m}{\mu'-U(m-1)}
+    \frac{m+1}{Um-\mu'}
\right]\,
e^{- \beta E_m(\mu')} = \frac{1}{2dJ}
\end{eqnarray}
with
\begin{equation}
\label{Z}
Z_0(\mu)
=
\sum_{m=0}^\infty
e^{- \beta E_m(\mu)}
\;,
\quad
E_m(\mu)
=
\frac{U}{2}
\,
m(m-1)
-
\mu \, m
\end{equation}
being the partition function without hopping.

\subsection{Pure case}

In the pure case we have $p(\epsilon)=\delta(\epsilon)$, so we get from (\ref{phb}) 
for the phase boundary~\cite{BV}
\begin{equation}
\label{phbpure}
\frac{2dJ}{Z_0(\mu)}
\sum_{m=0}^\infty
\left[
    \frac{m}{\mu-U(m-1)}
    +
    \frac{m+1}{Um-\mu}
\right]\,
e^{- \beta E_m(\mu)}
=
1
\;.
\end{equation}
In the zero-temperature limit this equation reduces to the well-known result
for the boundary between the superfluid and Mott-insulator~\cite{Sachdev,Oosten}
\begin{equation}
\label{Jd0pure}
2dJ
=
\frac
{
 \left[
     \mu - U(n-1)
 \right]
 \left[
     U n - \mu
 \right]
}
{\mu+U}
\;.
\end{equation}
Here $n$ denotes the positive integer at which $E_m(\mu)$ is minimal with respect to $m$.
This fixes $n$ as the smallest integer larger than or equal to $\mu/U$.

\subsection{Disorder with homogeneous distribution}

In the following we choose for simplicity
a homogeneous disorder distribution in the interval $\epsilon\in[-\Delta/2,\Delta/2]$
\begin{equation}
\label{hom}
p(\epsilon)
=
\frac{1}{\Delta}
\Big[
    \Theta
    \left(
        \epsilon+\Delta/2
    \right)
    -
    \Theta
    \left(
        \epsilon-\Delta/2
    \right)
\Big] \, ,
\end{equation}
so we have for the phase boundary~(\ref{phb})
\begin{eqnarray}
\label{pbh}
&&
\frac{2dJ}{\Delta}
\int_{\mu-\Delta/2}^{\mu+\Delta/2}
\frac{d\mu'}{Z_0(\mu')}
\sum_{m=0}^\infty
\left[
    \frac{m}{\mu'-U(m-1)}
    +
    \frac{m+1}{Um-\mu'}
\right]
e^{- \beta E_m(\mu')}
=
1
\;.
\end{eqnarray}

\begin{figure}[tb]
\centering

\psfrag{m}[l]{$\mu/U$}
\psfrag{j}[r]{$2dJ/U$}
\psfrag{M}[c]{MI}

\psfrag{a}[c]{\hspace{2mm}(a)}
\psfrag{t}[c]{$T = 0$}
\psfrag{d}[c]{$\Delta/U = 0.5$}

\psfrag{S}[c]{SF}
\includegraphics[width=6cm]{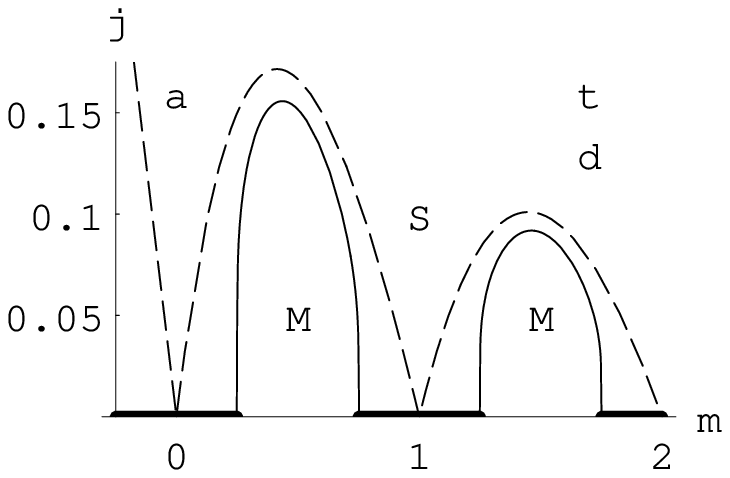}
\hspace{3mm}
\psfrag{c}[c]{\hspace{6mm}(c)}
\psfrag{S}[c]{\hspace{-15mm}SF}
\psfrag{G}[c]{ BG}
\psfrag{t}[c]{$kT/U = 0.01$}
\psfrag{d}[c]{$\Delta/U = 1$}
  \includegraphics[width=6cm]{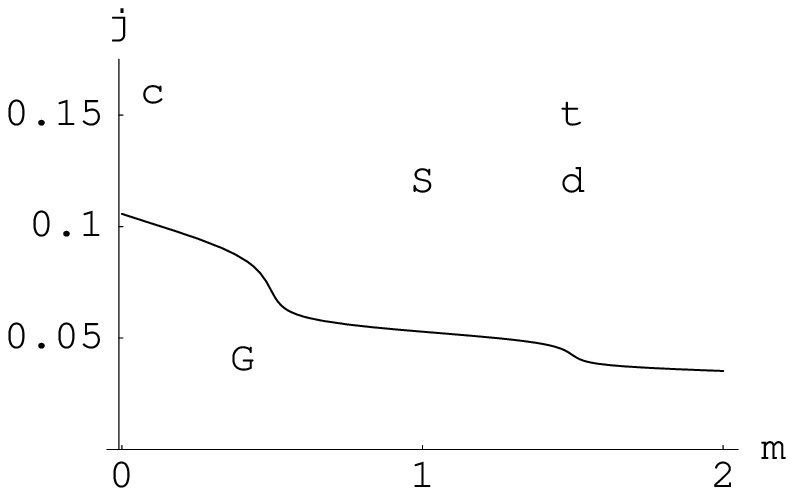}

\psfrag{b}[c]{\hspace{6mm}(b)}
\psfrag{S}[c]{SF}
\psfrag{G}[c]{ BG}
\psfrag{t}[c]{$kT/U = 0.01$}
\psfrag{d}[c]{$\Delta/U = 0.5$}
  \includegraphics[width=6cm]{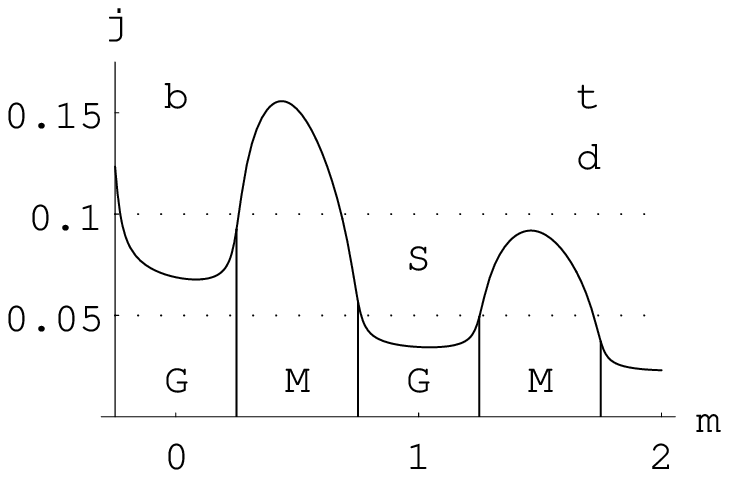}
\hspace{3mm}
\psfrag{e}[c]{\hspace{4mm}(d)}
\psfrag{G}[c]{\hspace{4mm}BG}
\psfrag{S}[c]{\hspace{-15mm}SF}
\psfrag{d}[c]{$\Delta/U = 1.5$}
  \includegraphics[width=6cm]{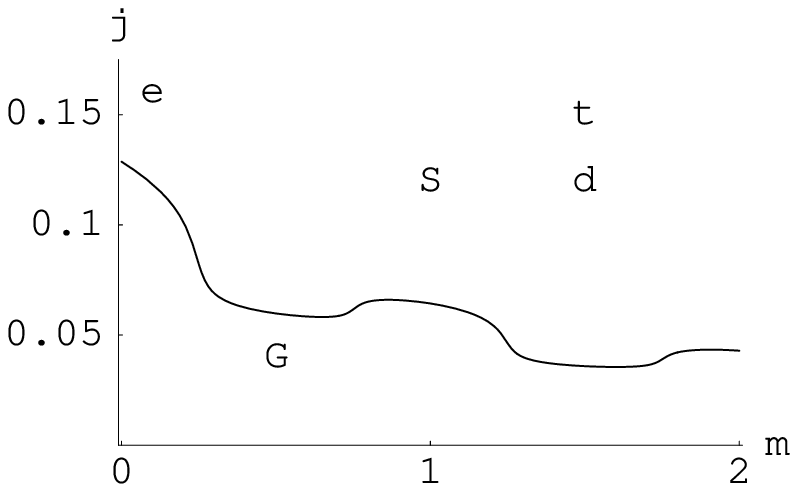}

\caption{Mean-field $(\mu,J)$ phase diagram for the homogeneous disorder
         distribution~(\ref{hom}).
         At $T=0$ (a), the Bose-glass phase exists only for $J=0$ (bold lines).
         The dashed line shows the boundary~(\ref{Jd0pure}) between the superfluid phase
         and the Mott-insulator phase in the pure case.
         At $T\ne0$ (b,c,d), the Bose glass phase exists at $J\ne0$ as well.
         If the disorder becomes strong (c,d), the Mott-insulator phase disappears.
         MI=Mott insilator with energy gap, BG=Bose glass with nonvanishing density
         of states at zero energy, SF=superfluid with nonvanishing superfluid density.
         The dotted lines in (b) indicate the two values of $J$ used to obtain the
         plots in Figs.~\ref{psi2},~\ref{comp},~\ref{mtd}.
        } 
\label{phd}
\end{figure}

We discuss first the special case $T=0$ and consider $\Delta<U$. 
It is assumed that 
$\mu\in[U(n-1),Un]$, $n=1,2,\dots$.
Eq.~(\ref{pbh}) then gives
\begin{equation}
\label{Jd0}
2dJ
=
\Delta
\left[
    n
    \ln
    \frac{\mu-U(n-1)+\Delta/2}{\mu-U(n-1)-\Delta/2}
    -
    (n+1)
    \ln
    \frac{Un-\mu-\Delta/2}{Un-\mu+\Delta/2}
\right]^{-1}
\;,
\end{equation}
if
$
\mu\in
[U(n-1)+\Delta/2,Un-\Delta/2]
$, otherwise $2dJ=0$.
In the limit $\Delta\to0$, Eq.~(\ref{Jd0}) reduces to (\ref{Jd0pure}).
The phase boundary at $T=0$ following from Eq.~(\ref{Jd0}) is shown in Fig.~\ref{phd}a
for typical parameter values.
It has a lobe structure and the size of the lobes decreases with increasing $\Delta$.
If $T=0$ but $\Delta>U$, we obtain $2dJ=0$ as the transition line for any value of $\mu$,
i.e., the lobes disappear and the superfluid phase appears as soon as 
$J$ is turned on~\cite{Fisher}.

\begin{figure}[tb]
\centering

\psfrag{m}[l]{$\mu/U$}
\psfrag{p}[r]{$\left|\psi_m\right|^2$}

\psfrag{a}[c]{(a)}
\psfrag{b}[c]{(b)}

\includegraphics[width=6cm]{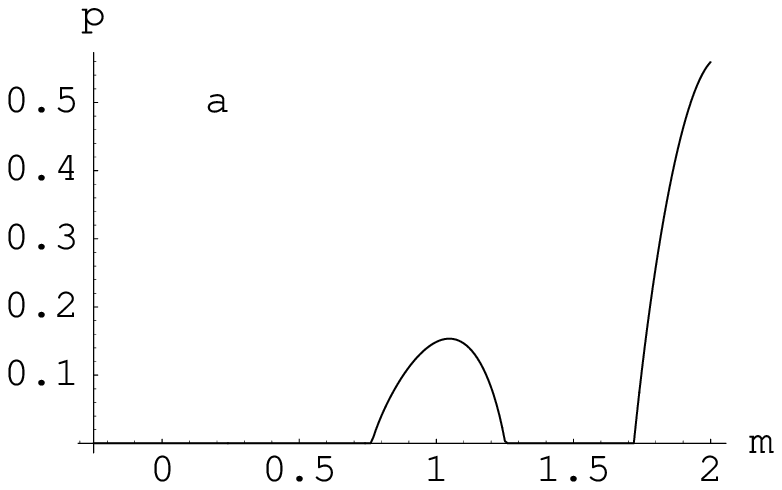}
\includegraphics[width=6cm]{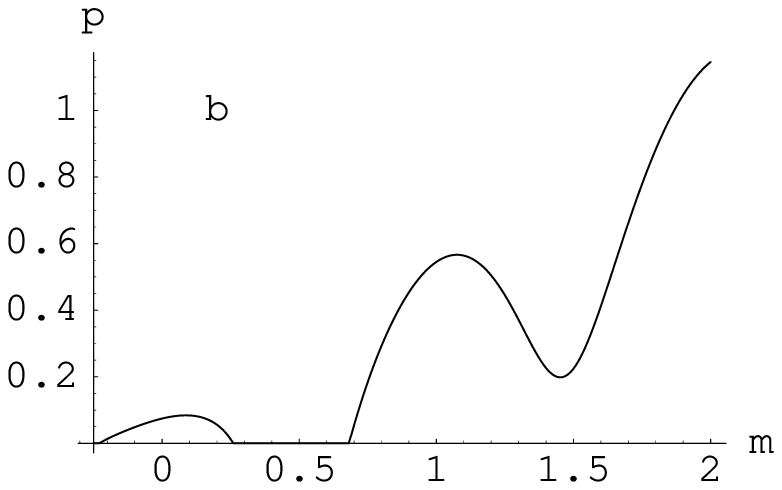}

\caption{Superfluid density for $kT/U=0.01$, $\Delta/U=0.5$, $2dJ/U=0.05$ (a), $0.1$ (b)
         (along the dotted lines in Fig.~\ref{phd}b).
        }
\label{psi2}
\end{figure}

In the case of non-zero temperature, Eq.~(\ref{pbh}) gives always non-vanishing
values of $2dJ$.
The boundary between the superfluid and non-superfluid phases for small $T$ 
and different values of $\Delta$ is shown in Figs.~\ref{phd}b,c,d.
The plots obtained from Eq.~(\ref{pbh}) and that obtained
by numerically diagonalizing the Hamiltonian~(\ref{hmf}) and minimizing
the free energy with respect to $\psi$ are indistinguishable.
If the temperature increases, the boundary between the superfluid and
non-superfluid phases goes upwards and the size of the non-superfluid region grows.
The presence of even a small temperature ($kT/U=0.01$) changes rather strongly
the phase boundary. This is in contrast to the pure case where values of $kT/U\sim 1$
are required in order to get a noticeable shift of the boundary between
the Mott-insulator and the superfluid phases~\cite{STO}.
The superfluid density $\left|\psi_m\right|^2$ obtained by the numerical diagonalization
for the same parameters as in Fig.~\ref{phd}b and the values of $J$ indicated
by dotted lines in Fig.~\ref{phd}b is plotted in Fig.~\ref{psi2}.

\section{Boundary between the Mott-insulator and Bose-glass phases}

In the Mott-insulator as well as in the Bose-glass phase the superfluid order parameter
$\psi=\psi_m$ vanishes, which implies that $J$ disappears from the mean-field 
Hamiltonian~(\ref{hmf}).
Therefore, the properties of these two phases do not depend on $J$ in the
mean-field approximation.
This is related to the fact that the two phases are localized and, therefore,
the dependence on $J$ should be weak. However, more accurate calculations
beyond the mean-field theory should give some dependence on $J$, in particular close
to the boundary with the superfluid phase.
The Mott-insulator phase is characterized by the gap in the excitation spectrum
and it has a finite superfluid susceptibility.
The Bose-glass phase has no gap and the superfluid susceptibility diverges.
All this is directly related to the properties of the Green's functions and
the density of states.

\subsection{Green's function}

The bosonic single-particle Green's function $G(t)$ is defined as~\cite{FW}
\begin{eqnarray}
\label{G}
&&
G(t)
=
-i
\left[
\Theta(t)
G_>(t)
+
\Theta(-t)
G_<(t)
\right]
\;,
\nonumber\\
&&
G_>(t)
=
\langle
     a(t)
     a^\dagger(0)
\rangle
\;,
G_<(t)
=
\langle
     a^\dagger(0)
     a(t)
\rangle
\;,
\end{eqnarray}
where $a(t)$ is the annihilation operator in the Heisenberg representation.
Straightforward calculations lead to the result
\begin{eqnarray}
\label{Ggs}
G_>(t) &=& \frac{1}{Z_0(\mu')} \sum_{m=0}^{\infty} (m+1)
\,\,e^{(\mu'-Um) it - \beta E_m(\mu')} \;,
\nonumber\\
G_<(t) &=&\frac{1}{Z_0(\mu')} \sum_{m=0}^{\infty} m \,\,
e^{[\mu'-U(m-1)] it - \beta E_m(\mu')} \;,
\end{eqnarray}
where $\mu'=\mu+\epsilon$ denotes the random local chemical potential.
One can easily show that the imaginary-time Green's functions satisfy the periodicity condition
$G_>(\tau+\beta) = G_<(\tau)$, where $\tau=it$ is the imaginary time.

At $T=0$ the imaginary-time Green's function takes the form
\begin{eqnarray}
\label{Gg0}
&&
\overline{G_>(\tau)} = \frac{(n_-+n_++1) (n_+-n_-) }{2 \tau\Delta}
+ \frac{1}{\tau\Delta} \Bigg[ (n_++1) \,
e^{- ( U n_+ - \mu - \Delta/2 ) \tau} \nonumber\\
&& - (n_-+1)\, e^{- (U n_- - \mu + \Delta/2) \tau }
- \frac{1}{2} \,(n_-+n_++3) (n_+-n_-) \,e^{- U \tau} \Bigg]
\;,
\end{eqnarray}
where $n_\pm$ is the smallest integer greater than or equal to
$(\mu \pm \Delta/2)/U$.
If $n_+=n_-$ which corresponds to 
$\mu\in[U(n-1) + \Delta/2 , U n - \Delta/2]$ for $\Delta<U$,
the first term in Eq.~(\ref{Gg0}) vanishes and the superfluid susceptibility
defined by the integral over the real-time Green's function
$\overline{\chi}=\int_0^\infty \overline{G_>(t)}dt$ is a finite quantity.
This means that we have not the Bose-glass phase, i.e., 
we are in the Mott-insulator phase (see Fig.~\ref{phd}a).
If $n_+ > n_-$ which corresponds to
$\mu\in[U(n-1) , U(n-1) + \Delta/2]\cup[U n - \Delta/2 , U n]$ for $\Delta<U$
or arbitrary $\mu$ for $\Delta>U$,
the first term in Eq.~(\ref{Gg0}) survives and renders
$\overline{\chi}$ divergent which is the distinguishing property of the Bose-glass phase.
In the case $\Delta>U$, the lobes in Fig.~\ref{phd}a disappear completely
which means that the Mott-insulator phase is destroyed by the disorder~\cite{Fisher}.

\begin{figure}[tb]
\centering

\psfrag{t}[l]{$\tau U$}
\psfrag{G}[l]{$\overline{G_>}$}

  \includegraphics[width=6cm]{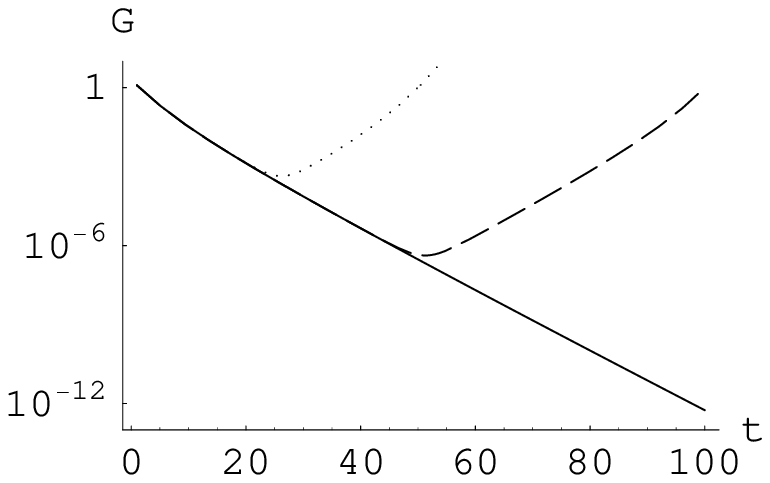}
  \hspace{5mm}
  \includegraphics[width=6cm]{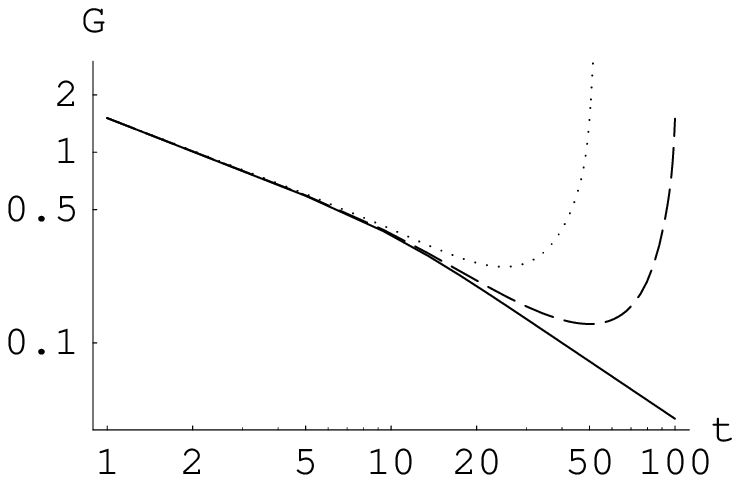}

(a)\hspace{6cm}(b)
\caption{Correlation 
function $\overline{G_>}$ for $\Delta/U=0.5$ in the Mott-insulator [$\mu/U=0.5$ (a)]
         and Bose-glass [$\mu/U=1$ (b)] phase. $T=0$ (solid lines), $kT/U=0.01$ (dashed lines),
         $kT/U=0.02$ (dotted lines).
         Note that the scale of $\overline{G_>}$ is logarithmic but the scale of
         $\tau$ is linear in (a) and logarithmic in (b), respectively.
        }
\label{green}
\end{figure}

At non-zero temperature, it is more difficult to analyze the structure of 
the disorder averaged Green's function.
Expanding Eq.~(\ref{Ggs}) for large but finite values of $\beta$ shows
that the Green's function has a similar structure as Eq.~(\ref{Gg0}) but the explicit
expressions become very long and we do not display them.
Typical $\tau$-dependences of $\overline{G_>}$ in the Mott-insulator as well as in the Bose-glass 
phase are shown for different temperatures in Fig.~\ref{green}.
Due to the different scales of $\tau$, Figs.~\ref{green}a and~\ref{green}b
indicate that the Mott-insulator and the Bose-glass phase are characterized
by an exponential and algebraic decay of $\overline{G_>(\tau)}$, respectively.
These analytical results agree qualitatively with Monte Carlo
simulations~\cite{WSGY,KR,KR2,LeeCha,Hitchcock}.

Since even at finite temperature $\overline{G_>(\tau)}$ decays only like $1/\tau$, 
the superfluid susceptibility still diverges logarithmically.
Vice versa, the divergent superfluid suceptibility $\overline{\chi}$ has
the consequence that the density of states for the single-particle excitations
at zero energy does not vanish in contrast
to the case of finite $\overline{\chi}$~\cite{Fisher}. In the next section we will see that
the density of states is easier to analyze at finite
temperature than the full Green's function.

\subsection{Density of states}

\begin{figure}[tb]
\centering

\psfrag{r}[l]{$\overline{\rho}U$}
\psfrag{e}[l]{$\omega/U$}
\psfrag{a}[c]{(a)}
\psfrag{b}[c]{(b)}
  \includegraphics[width=6cm]{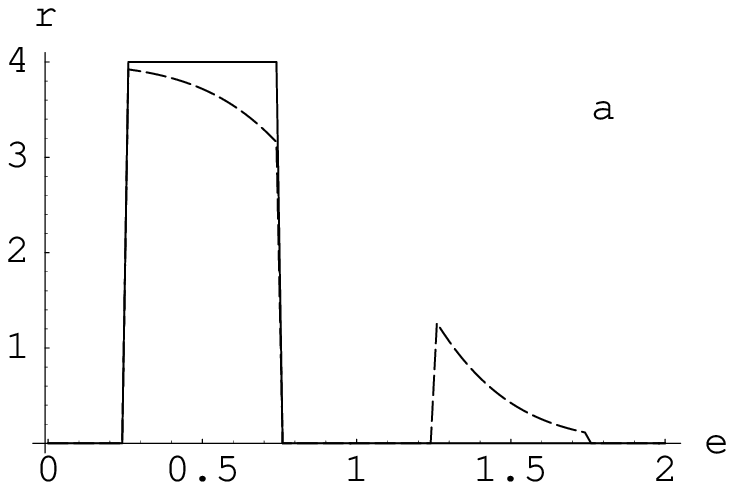}
  \includegraphics[width=6cm]{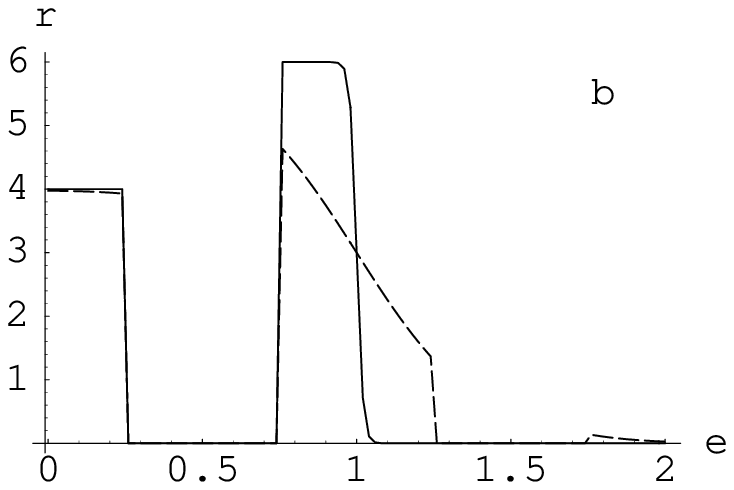}

\caption{(a) Density of states in the Mott-insulator phase for $\Delta/U=0.5$, $\mu/U=0.5$,
         $k T/U=0.01$ (solid line), $0.2$ (dashed line).
         (b) Density of states in the Bose-glass phase for $\Delta/U=0.5$, $\mu/U=1$,
         $k T/U=0.01$ (solid line), $0.2$ (dashed line).
        }
\label{ds}
\end{figure}

\begin{figure}[tb]
\centering

\psfrag{b}[l]{$\beta U$}
\psfrag{r}[l]{$\overline{\rho(0)}U$}

  \includegraphics[width=6cm]{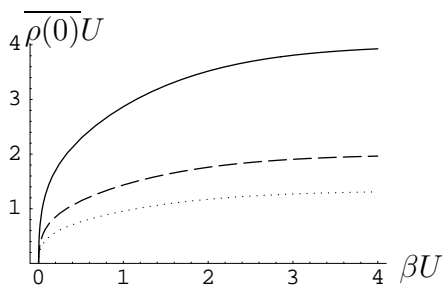}

\caption{Density of states at zero energy in the Bose-glass phase for
         $\mu/U=1$, $\Delta/U=0.5$ (solid line), $1$ (dashed line), $1.5$ (dotted line).
        }
\label{ds0}
\end{figure}

The density of states for the single-particle excitations can be determined in terms of
the Fourier transformed single-particle Green's function
$
\tilde G(\omega)
=
\int_{-\infty}^{+\infty}
d\,t
\exp
\left(
    i \omega t
\right)
G(t)
$
as
$
 \rho(\omega,\mu)
 =
 -
 \frac{1}{\pi}\,
 {\rm Im}\,
 \tilde G(\omega)
$~\cite{AGD}. The Fourier transformation 
of Eqs.~(\ref{G}), (\ref{Ggs}) gives the density of states 
for the pure case
\begin{eqnarray}
\rho(\omega,\mu)
&=&
\frac{1}{Z_0(\mu)}
\sum_{m=0}^\infty
e^{- \beta E_m(\mu)} \nonumber \\
&& \times
\Big[
    m\,
    \delta
    \left(
     \omega + \mu - U (m-1)
    \right)
    +
    (m+1)
    \delta
    \left(
       \omega + \mu - U m
    \right)
\Big]
\;.
\end{eqnarray}
The two $\delta$-functions correspond to the hole and particle excitations, respectively.
After the disorder averaging we obtain
\begin{eqnarray}
\overline{\rho(\omega,\mu)} &=&
\sum_{m=0}^\infty
\frac
{(m+1)\,p(Um-\mu-\omega)}
{Z_0(Um-\omega)} 
\nonumber \\
&& \times
\Big[
e^{- \beta E_{m+1}(Um-\omega)} 
+ e^{- \beta E_{m}(Um-\omega)}
\Big]
\,.
\end{eqnarray}
This disorder averaged density of states is plotted in Fig.~\ref{ds}a for the Mott-insulator
phase with the energy gap $\omega_g = U n - \Delta/2 - \mu$ and in   Fig.~\ref{ds}b
for the Bose-glass phase.
Since $E_{m+1}(Um)=E_m(Um)$, we get finally
\begin{equation}
\label{r0}
\overline{\rho(0,\mu)}
=
2 \sum_{m=0}^\infty
\frac
{(m+1)\,p(Um-\mu)}
{Z_0(Um)}
\,\,e^{- \beta E_{m}(Um)}
\;.
\end{equation}
For the homogeneous disorder distribution (\ref{hom}) the summation in Eq.~(\ref{r0})
is restricted by $m=n_-,\dots,n_+'$, where $n_-$ is the smallest integer greater than or equal to
$(\mu - \Delta/2)/U$ and $n_+'$ is the greatest integer less than or equal to
$(\mu + \Delta/2)/U$.
If $\Delta<U$, Eq.~(\ref{r0}) takes the form
\begin{equation}
\label{rt}
\overline{\rho(0,\mu)}
=
\left\{
\begin{array}{ll}
{\displaystyle \frac{2 n}{\Delta Z_0(U(n-1))} }
\,\,e^{ - \beta E_{n-1}(U(n-1))}
&
\mu\in{\cal G}_1
\\
0
&
\mu
\in
{\cal M}
\\
{\displaystyle \frac{2 (n+1)}{\Delta Z_0(Un)}}
\,\,e^{- \beta E_n(Un)}
&
\mu
\in
{\cal G}_2 \, ,
\end{array}
\right.
\end{equation}
where we defined $n=n(\mu)$ as the smallest integer larger than or equal $\mu / U$ and where
$
{\cal G}_1
=
\left[
    U(n-1) , U(n-1) + \Delta/2
\right]
$,
$
{\cal G}_2
=
\left[
    U n - \Delta/2 , U n
\right]
$,
$
{\cal M}
=
\left[
    U(n-1) + \Delta/2 , U n - \Delta/2
\right]
$.
The temperature dependence of $\overline{\rho(0,\mu)}$ is plotted in Fig.~\ref{ds0}.
In the limit $T\to0$,
Eq.~(\ref{rt}) reduces to
\begin{equation}
\label{rh0}
\overline{\rho(0,\mu)}
=
\left\{
\begin{array}{ll}
n/\Delta
&
\mu\in
{\cal G}_1
\\
0
&
\mu
\in
{\cal M}
\\
(n+1)/\Delta
&
\mu
\in
{\cal G}_2 \, .
\end{array}
\right.
\end{equation}
Simple analytical expressions (\ref{rt}), (\ref{rh0}) show that
the lines $\mu=U(n-1)+\Delta/2$ and $\mu=Un-\Delta/2$ determine the boundaries
between the Mott-insulator and Bose-glass phases at arbitrary temperature (see Fig.~\ref{phd}).
In the case $\Delta>U$ of strong disorder, $\overline{\rho(0,\mu)}$ does not vanish
for finite temperature and the Mott-insulator phase does not exist.

%

\section{Compressibility}

The compressibility of the system is defined as
$
\overline{\kappa(\mu)}
=
-
\partial^2 \overline{F(\mu)}
/
\partial \mu^2
$,
where
$
\overline{F(\mu)}
=-
\overline{\ln Z(\mu)}/\beta
$.
In a non-superfluid phase $Z(\mu)=Z_0(\mu)$ is given by Eq.~(\ref{Z}).
Partial integration over $\epsilon$ gives
\begin{equation}
\overline{\kappa(\mu)}
=
- \frac{1}{\beta}
\frac{\partial}{\partial\mu}
\int_{-\infty}^{+\infty}
\ln Z_0(\mu+\epsilon)
\frac{dp(\epsilon)}{d\epsilon}
d\epsilon
\;.
\end{equation}
For the homogeneous distribution (\ref{hom}) we get
\begin{equation}
\label{kappa}
\overline{\kappa(\mu)}
=
\frac{1}{\Delta}
\Big[
    N
    \left(
        \mu + \Delta/2
    \right)
    -
    N
    \left(
        \mu - \Delta/2
    \right)
\Big]
\;,
\end{equation}
where
\begin{equation}
N(\mu)
=
\langle
a^\dagger
a
\rangle
=
\frac{1}{Z_0(\mu)}
\sum_{m=0}^{\infty}
m
\,e^{- \beta E_m(\mu)}
\end{equation}
is the mean particle number per lattice site in the pure case.
For the Bose-glass phase, the compressibility~(\ref{kappa}) does not vanish.
One can easily show that
\begin{eqnarray}
\lim_{\Delta\to0}
\overline{\kappa(\mu)}
&=&
\kappa(\mu)
=
\beta
\left[
    \langle
    (a^\dagger a)^2
    \rangle
    -
    \langle
    a^\dagger a
    \rangle^2
\right]
\nonumber\\
&
=&
\beta
\left[
\frac{1}{Z_0(\mu)}
\sum_{m=0}^{\infty}
m^2
\,e^{- \beta E_m(\mu)}
-
N^2(\mu)
\right]
\;.
\end{eqnarray}
If $\beta U \gg 1$, the compressibility~(\ref{kappa}) expanded for small temperatures
has the form
\begin{equation}
\overline{\kappa(\mu)}
\approx
\frac{n_+ - n_-}{\Delta}
+
\alpha\, e^{-\beta\delta}
\;,
\end{equation}
where
$\delta(\mu)=E_n-{\rm min}(E_{n-1},E_{n+1})$
is the energy difference between
the first excited state and the ground state in the pure case (cf.~(\ref{Z})), 
and $\alpha$ is some
finite constant. This equation shows that 
the Mott-insulator phase, which occurs for $n_+=n_-$, has an  exponentially small
compressibility at non-zero temperature, 
in contrast to the Bose-glass phase.

\begin{figure}[tb]
\centering

\psfrag{m}[l]{$\mu/U$}
\psfrag{k}[l]{$\overline{\kappa}\, U$}

  \includegraphics[width=6cm]{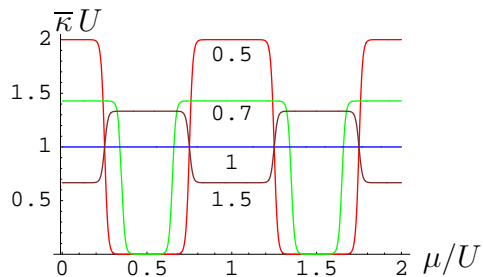}

\caption{Compressibility for $k T/U=0.01$, $J=0$ and the values of $\Delta/U$ given below the lines
         worked out according to Eq.~(\ref{kappa}).}
\label{comp0}
\end{figure}

\begin{figure}[tb]
\centering

\psfrag{m}[l]{$\mu/U$}
\psfrag{k}[l]{$\overline{\kappa}\, U$}

\psfrag{a}[c]{(a)}
\psfrag{b}[c]{(b)}
  
\psfrag{G}[c]{\hspace{2.5mm}BG}
\psfrag{S}[c]{SF}
\psfrag{M}[c]{MI}
  \includegraphics[width=6cm]{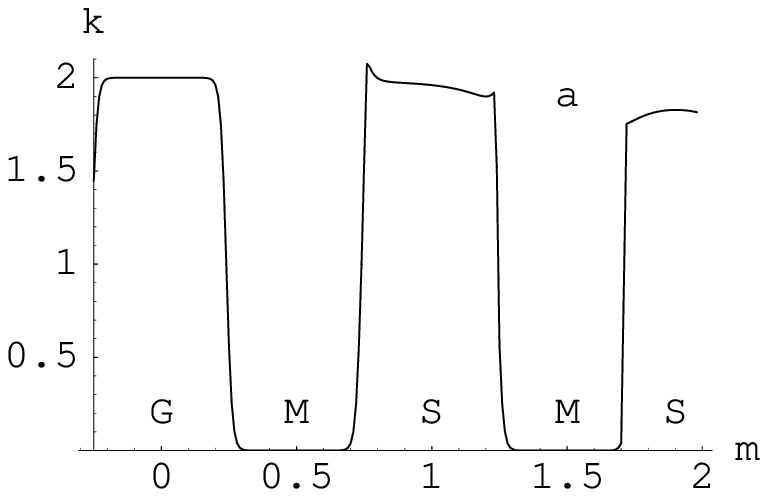}
  \hspace{5mm}
  \includegraphics[width=6cm]{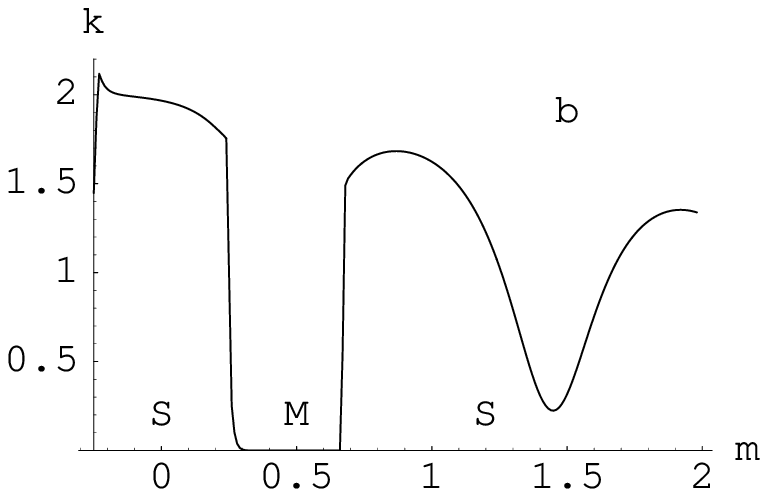}

\caption{Compressibility for $k T/ U=0.01$, $\Delta/U=0.5$, $2dJ/U=0.05$ (a), $2dJ/U=0.1$ (b)
         (along the dotted lines in Fig.~\ref{phd}b).
         The results are obtained by numerically diagonalizing the Hamiltonian~(\ref{hmf})
         and calculating the second derivative of the free 
energy with respect to $\mu$
         at $\psi=\psi_m$.}
\label{comp}
\end{figure}

The dependence of the compressibility on $\mu$ for small temperature
is shown in Figs.~\ref{comp0},\ref{comp}. Since the compressibility does not vanish at non-zero
temperature and is a continuous function of the system parameters, it can not be
used as a criterion to distinguish between different phases. Thus, we deduce
that the transitions are
better defined in terms of the superfluid order parameter $\psi$ and the density
of states.

%

\section{Conclusions}

The Mott-insulator phase and the Bose-glass phase at vanishing temperature 
can be defined either by their thermodynamic properties or by the spectral 
properties of their low-lying excitations. Both characterizations are, of course, 
closely  related. Both phases are non-superfluid, i.e., the corresponding Goldstone 
modes, the phonons, are absent. In the case of the Mott-insulator phase the spectral 
characterization by an energy gap implies a vanishing compressibility and vanishing 
particle-number fluctuations at $T=0$. In the Bose-glass phase the non-vanishing density 
of states at zero energy implies a non-vanishing compressibility. These features 
allow a sharp distinction between the two phases at zero temperature.

\begin{figure}[tb]
\centering

\psfrag{m}[l]{$\mu/U$}
\psfrag{t}[l]{$kT/U$}
\psfrag{M}[c]{MI}
\psfrag{S}[c]{SF}
\psfrag{G}[c]{BG}

\psfrag{a}[c]{(a)}
\includegraphics[width=6cm]{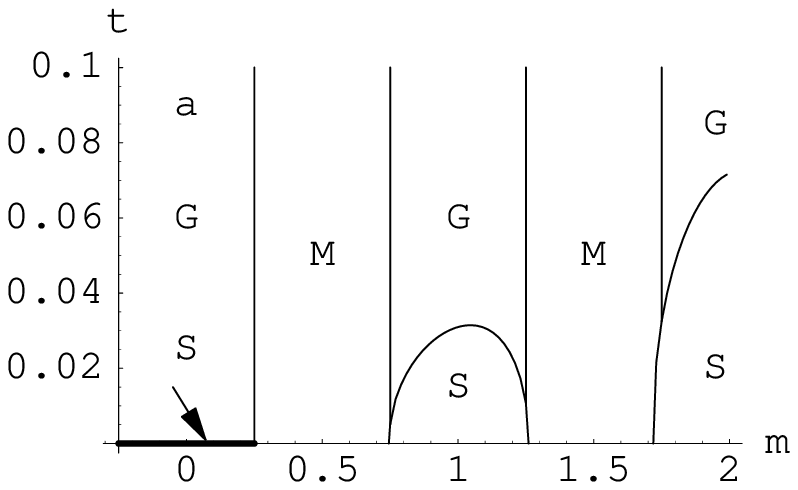} 
\hspace*{3mm}
\psfrag{b}[c]{(b)}
\includegraphics[width=6cm]{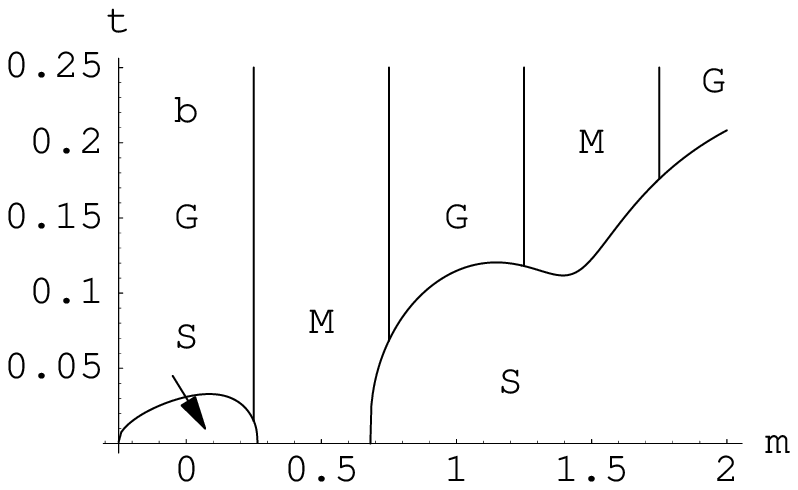}

\caption{Phase diagram in $(\mu,T)$-plane corresponding to the two dotted lines in
Fig.~\ref{phd}b, with the tunneling amplitude $2dJ/U=0.05$~(a), $0.1$~(b).}
\label{mtd}
\end{figure}

However, at non-vanishing temperatures, the characterization of the Mott-insulator and 
Bose-glass phases by their thermodynamic properties is no longer sharp -- 
the Mott-insulator phase has an exponentially small but finite compressibility which
corresponds to non-vanishing fluctuations of the particle number
density. Still, as we have pointed out 
in this paper, the characteristic spectral features remain present also at 
$T>0$ and can therefore be used for a sharp definition and distinction 
between these low temperature phases. We employed this possibility to 
calculate finite-temperature phase diagrams within a Bose-Hubbard model with on-site 
disorder within the mean-field approximation. For experiments with optical lattices 
it is usually easiest to change system parameters like the tunnelling amplitude 
$J$ at fixed temperature, i.e., phase diagrams of the format of Fig.~\ref{phd},  
where system parameters are used as variables, are most natural from this point 
of view. However, from a thermodynamic point of view, it may be more natural to give 
the phase diagram in the ($\mu,T$)-plane. This is done in Fig.~\ref{mtd} for the two 
values of $J$ marked in Fig.~\ref{phd}b as dotted lines. On the high temperature 
side, the phase diagram of Fig.~\ref{mtd} is incomplete, because there the 
transition to the normal gas phase must occur, which we have not considered in the present work.
For both the Mott-insulator and the Bose-glass phase this transition would be sharp,
if the energy gap or the finite density of states would start to appear suddenly
at a critical temperature. Alternatively, the transition could also take the form
of a smooth crossover. For the Mott-insulator phase the crossover could occur at the
temperature $k T \approx \omega_g$, where $\omega_g$ is the energy-gap for thermal 
excitations \cite{SKPR,STO}. 
For the Bose-glass phase this would happen at the temperature where the density
of states starts to be dominated by the normal Bose gas.

In our calculations, we find direct transitions from the superfluid phase
either to the Mott-insulator or to the Bose-glas phase under variation of $J$,
depending on the values of $\mu$ and the disorder strength $\Delta$ (Fig.~\ref{phd}).
The character of the transition from the superfluid to the Bose-glas phase
is in agreement with Monte Carlo calculations.
The direct transition from the superfluid to the Mott-insulator phase in the case
of weak disorder $\Delta<U$ is also
in agreement with the results obtained by the path-integral Monte Carlo
techniques~\cite{Krauth}
and by the Monte Carlo simulations based on the $J$-current model,
which neglects amplitude fluctuations of the bosonic quantum fields,
for $\mu/U=0.5$~\cite{KR,KR2,Kim,LeeCha} and for some finite
interval of $\mu$ near the tip of the Mott-insulator lobe~\cite{Park}.
However, more recent Monte Carlo investigations of the $J$-current model~\cite{PS},
still neglecting the amplitude fluctuations,
show that the superfluid and the Mott-insulator regions on the phase diagram are
separated by a narrow region occupied by the Bose-glas phase if the number of
lattice sites is large enough. Thus, different Monte Carlo techniques
give different results on the character of the superfluid -- Mott-insulator
transition in the presence of disorder.
Other methods lead also to mutually conflicting results concerning this 
point~\cite{review}. Here we have presented the results of the mean-field approach.

Experimentally, the Bose-glass phase may not be easy to identify with ultracold 
atoms in a suitably disordered lattice. In fact, it might best be identifiable 
indirectly by the absence of properties which are present in the competing phases 
for $T\to 0$, like the absence of a macroscopic wave function and the absence 
of an energy gap or of incompressibility~\cite{NEU}. 
The finite density of states at $\omega \to 0$ would show up in a 
specific heat proportional to $T$ for $T \to 0$ and in a logarithmically 
diverging susceptibility $\overline{\chi}$.
It would certainly be of great interest if a way could be found to measure 
$\overline{\rho (0)}$ directly. 

\ack

This work was supported by the SFB/TR 12 of the German Research Foundation (DFG).

\section*{References}


\begin{thebibliography}{99}

\bibitem{Bloch1}
Greiner M, Mandel O, Esslinger T, H{\"a}nsch T W and Bloch I 
2002
{\it Nature} {\bf 415} 39 

\bibitem{Zwerger}
Zwerger W 
2004 
{\it Advances in Solid State Physics} {\bf 44} 277 

\bibitem{Esslinger}
St{\"o}ferle T, Moritz H, Schori C, K{\"o}hl M and Esslinger T
2004
\PRL {\bf 92} 130403 

\bibitem{Zoller}
Jaksch D and Zoller P 
2005 
\AP {\bf 315} 52

\bibitem{Bloch2}
Bloch I
2005
{\it Nature Physics} {\bf 1} 23

\bibitem{Review}
Morsch O and Oberthaler M 
2006
\RMP {\bf 78} 179 

\bibitem{Inguscio1}
Lye J E, Fallani L, Modugno M, Wiersma D S and Inguscio M 
2005
\PRL {\bf 95} 070401

\bibitem{Inguscio2}
Fort C, Fallani L, Guarrera V, Lye J E, Modugno M, Wiersma D S and Inguscio M 
2005
\PRL {\bf 95} 170410 

\bibitem{Aspect}
Cl{\'e}ment D, Var{\'o}n A F, Hugbart M, Retter J A, Bouyer P, Sanchez-Palencia L, 
Gangardt D M, Shlyapnikov G V and Aspect A 
2005
\PRL {\bf 95} 170409 

\bibitem{Vignolo}
Vignolo P, Akdeniz Z and Tosi M P
2003
\JPB {\bf 36} 4535

\bibitem{Castin}
Gavish U and Castin Y 
2005
\PRL {\bf 95} 020401 

\bibitem{Sengstock}
Ospelkaus S, Ospelkaus C, Wille O, Succo M, Ernst P, Sengstock K and Bongs K
2006
\PRL {\bf 96} 180403

\bibitem{Schmiedmayer}
Kruger P, Andersson L M , Wildermuth S, Hofferberth S, Haller E, 
Aigner S, Groth S, Bar-Joseph~I and Schmiedmayer~J
2005
eprint: {\tt cond-mat/0504686}

\bibitem{GWSS05}
Gimperlein~H, Wessel~S, Schmiedmayer~J and Santos~L
2005
\PRL {\bf 95} 170401

\bibitem{GWSS06}
Gimperlein~H, Wessel~S, Schmiedmayer~J and Santos~L
2006
{\it Appl.~Phys.~B} {\bf 82} 217

\bibitem{Lew03}
Damski B, Zakrzewski J, Santos L, Zoller P and Lewenstein M
2003
\PRL {\bf 91} 080403

\bibitem{Santos05}
Sanchez-Palencia L and Santos L
2005
\PR A {\bf 72}, 053607

\bibitem{Schulte}
Schulte T, Drenkelforth S, Kruse J, Ertmer W, Arlt J, Sacha K, Zakrzewski J and Lewenstein M
2005
\PRL {\bf 95}, 170411

\bibitem{NEU}
Fallani L, Lye J E, Guarrera V, Fort C and Inguscio M
2006
eprint: {\tt cond-mat/0603655}

\bibitem{Kirkpatrick}
Belitz D and Kirkpatrick T R 
1994
\RMP {\bf 66} 261 

\bibitem{BECBCS1}
Bartenstein M, Altmeyer A, Riedl S, Jochim S, Chin C, Denschlag J H and Grimm R 
2004
\PRL {\bf 92} 120401 

\bibitem{BECBCS2}
Zwierlein M W, Stan C A, Schunck C H, Raupach S M F, Kerman A J and Ketterle W
2004
\PRL {\bf 92} 120403 

\bibitem{Reppy1}
Crooker B C, Hebral B, Smith E N, Takano Y and Reppy J D 
1983
\PRL {\bf 51} 666 

\bibitem{Reppy2}
Chan M H W, Blum K I, Murphy S Q, Wong G K S and Reppy J D 
1988
\PRL {\bf 61} 1950

\bibitem{Reppy3}
Reppy J D 
1992
{\it J. Low Temp. Phys.} {\bf 87} 205

\bibitem{Anderson}
Hertz J A, Fleishman L and Anderson P W 
1979
\PRL {\bf 43} 942 

\bibitem{Ordnungsparameter}
Graham R and Pelster A
2005
eprint: {\tt cond-mat/0508306}

\bibitem{AGD}
Abrikosov A A, Gor'kov L P and Dzyaloshinskii I Ye
1965 
{\it Quantum Field Theoretical Methods in Statistical Physics}
(Pergamon Press)

\bibitem{Gunn}
Lee D K K and Gunn J M F
1990
J. Phys. {\bf 2} 7753

\bibitem{Zhang1}
Nisamaneephong P, Zhang L and Ma M 
1993
\PRL {\bf 71} 3830

\bibitem{Zhang2}
Ma M, Nisamaneephong P and Zhang L 
1993 
{\it J. Low Temp. Phys.} {\bf 93} 957

\bibitem{Huang}
Huang K and Meng H F 
1992
\PRL {\bf 69} 644 

\bibitem{Giorgini}
Giorgini S, Pitaevskii L, and Stringari S 
1994
\PR B {\bf 49} 12938 

\bibitem{SR}
Singh K G and Rokhsar D S
1994
\PR B {\bf 49} 9013 

\bibitem{Kobayashi}
Koboyashi M and Tsubota M 2002
\PR B {\bf 66} 174516 

\bibitem{Vinokur}
Lopatin A V and Vinokur V M
2002
\PRL {\bf 88} 235503

\bibitem{Lee}
Ma M, Halperin B I and Lee P A 
1986
\PR B {\bf 34} 3136 

\bibitem{Fisher}
Fisher M P A, Weichman P B, Grinstein G and Fisher D S
1989 
\PR B {\bf 40} 546

\bibitem{Pazmandi}
P\'azm\'andi F, Zim\'anyi G and Scalettar R
1995
\PRL {\bf 75} 1356

\bibitem{Sing}
Singh K G and Roksar D S 
1992
\PR B {\bf 46} 3002

\bibitem{Pai}
Pai R V, Pandit R, Krishnamurthy H R and Ramasesha S
1996
\PRL {\bf 76} 2937

\bibitem{Svistunov}
Svistunov B V 
1996
\PR B {\bf 54} 16131

\bibitem{PZ}
P\'azm\'andi F and Zim\'anyi G
1998
\PR B {\bf 57} 5044

\bibitem{Rapsch}
Rapsch S, Schollw\"ock U and Zwerger W
1999
{\it Europhys. Lett.} {\bf 46} 559

\bibitem{Krauth}
Krauth W, Trivedi N and Ceperley D
1991
\PRL {\bf 67} 2307

\bibitem{Batrouni}
Batrouni G G and Scalettar R T
1992
\PR B {\bf 46} 9051

\bibitem{WSGY}
Wallin M, S{\o}rensen E S, Girvin S M and Young A P
1994 
\PR B {\bf 49} 12115

\bibitem{KR}
Kisker J and Rieger H,
1997 
{\it Physica} A {\bf 246} 348

\bibitem{KR2}
Kisker J and Rieger H,
1997
\PR B {\bf 55} R11981

\bibitem{Park}
Park S Y, Lee J W, Cha M C, Choi M Y, Kim B J and Kim D
1999
\PR B {\bf 59} 8420

\bibitem{Kim}
Lee J W, Cha M C and Kim D
2001
\PRL {\bf 87} 247006

\bibitem{LeeCha}
Lee J W and Cha M C
2004
\PR B {\bf 70} 052513

\bibitem{PS}
Prokof'ev N and Svistunov B
2004
\PRL {\bf 92} 015703

\bibitem{Hitchcock}
Hitchcock P and S{\o}rensen E S
2006
\PR B {\bf 73} 174523

\bibitem{Runge}
Runge K J
1992
\PR B {\bf 45} 13136

\bibitem{Davidson}
Pugatch R, Bar-gill N, Katz N, Rowen E and Davidson N 
2006
eprint: {\tt cond-mat/0603571}

\bibitem{SKPR}
Sheshadri K, Krishnamurthy H R, Pandit R and Ramakrishnan T V
1993 
{\it Europhys. Lett.} {\bf 22} 257

\bibitem{Sachdev}
Sachdev S
2001 
{\it Quantum phase transitions}
(Cambridge: Cambridge University Press)

\bibitem{Oosten}
van Oosten D, van der Straten P and Stoof H T C
2001 
\PR A {\bf 63} 053601

\bibitem{RB03}
Roth R and Burnett K
2003
\PR A {\bf 67} 031602(R)

\bibitem{BV}
Buonsante P and Vezzani A
2004 
\PR A {\bf 70} 033608

\bibitem{STO}
Dickerscheid D B M, van Oosten D, Denteneer P J H and Stoof H T C
2003
\PR A {\bf 68} 043623 

\bibitem{FW}
Fetter A L and Walecka J D
1971 
{\it Quantum Theory of Many-Particle Systems}
(New York: McGraw-Hill Book Company)

\bibitem{review}
Lewenstein M, Sanpera A, Ahufinger V, Damski B, Sen De A and Sen U
2006
eprint: {\tt cond-mat/0606771}

\end{thebibliography}
\end{document}